\documentclass[onecolumn,showpacs,preprintnumbers,amsmath,amssymb,superscriptaddress,nofootinbib]{revtex4}
\usepackage{graphicx}

\begin{document}

\title{Alpha decay rate enhancement in metals: An unlikely scenario}

\author{Nikolaj Thomas Zinner}\email{zinner@phys.au.dk}
\affiliation{Institute for Physics and Astronomy, University of {\AA}rhus,
DK-8000 {\AA}rhus C, Denmark}
\affiliation{Gesellschaft f\"ur Schwerionenforschung GSI, Darmstadt, Germany}

\date{\today}

\begin{abstract}
It has been recently suggested that one might drastically 
shorten the alpha lifetime
of nuclear waste products, if these are embedded in metals at low temperatures.
Using  quantum mechanical
tunneling arguments, we show that such an effect is likely to be very small, if present at all.
\end{abstract}


\maketitle

\section{Introduction}
In a recent publication, Kettner \emph{et. al.} \cite{Kettner1} have 
suggested that one could speed up
alpha decay of transuranic nuclear waste material by embedding it 
in metals at low temperature. This would 
have a huge impact on waste disposal management and quite likely initiate 
a completely renewed public 
debate on the subject of nuclear energy. The proposed idea is that the electron screening provided by 
the metal 
will lower  
the alpha-decay lifetime. Perturbations to nuclear alpha decay energies and barriers is a basic 
problem that was 
clarified and solved many years ago \cite{Emery72}. The influence 
of
an electron plasma environment on decay rates has also been carefully studied many times before and, 
as the present paper 
will demonstrate, the recently
proposed effect
is unlikely from the viewpoint of standard screening models.

\section{Tunneling and Screening}
\begin{figure}[ht!]
\centering
\includegraphics[width=0.95\textwidth,clip=true]{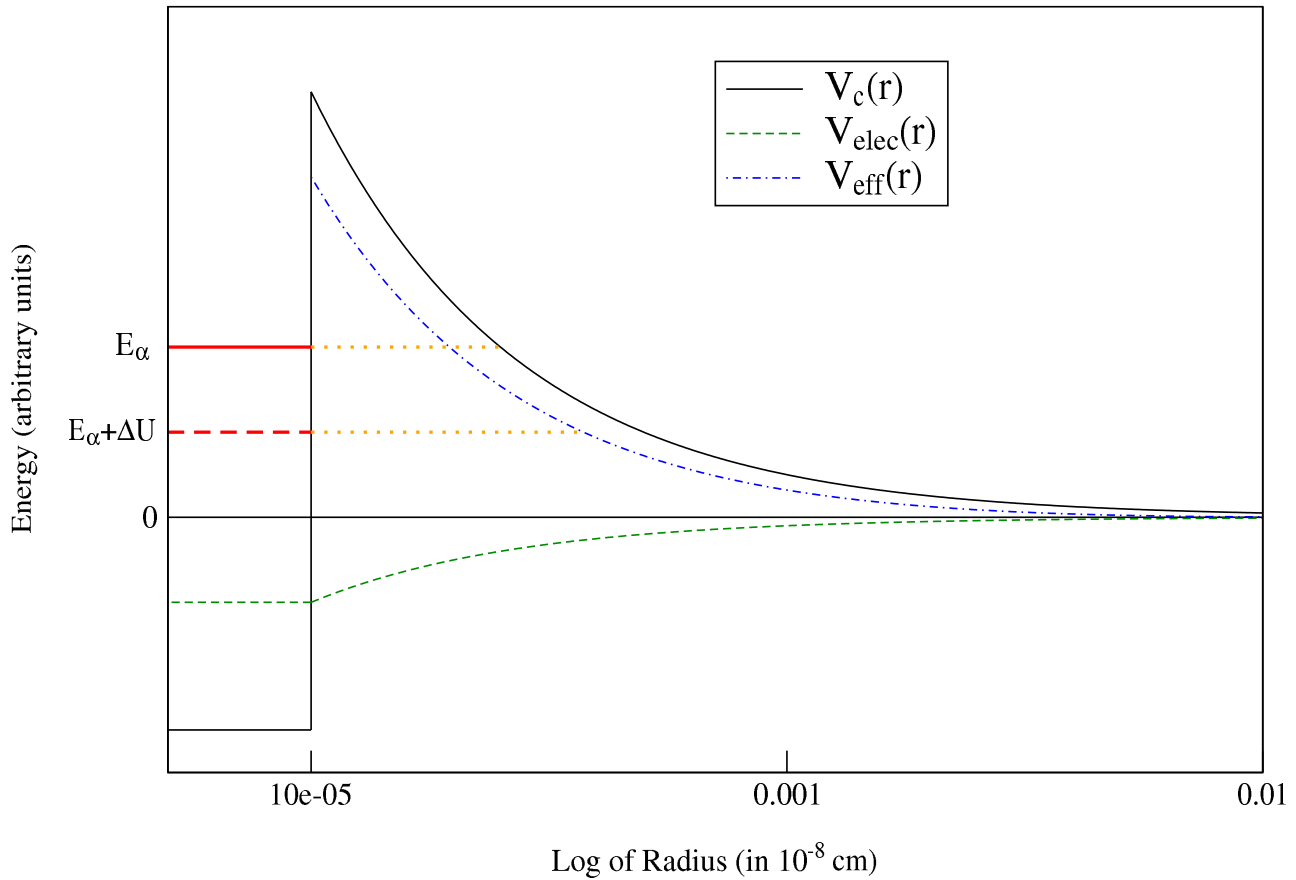}
\caption{Figure showing the basics of alpha tunneling. Shown are the pure coulomb barrier (full), the 
potential from 
the electron screening (dashed) and the effective screened potential (dash-dotted). The $\alpha$ particle 
energy with 
and without 
screening 
correction are shown inside the nuclear potential well. Also show are the 
tunneling paths. Notice that the radius is plotted on a log scale.} 
\label{fig:ill} 
\end{figure}

The quantum mechanics of tunnelling is described in virtually all standard textbooks (see for instance 
\cite{Griffiths1}). The case of alpha decay 
goes back to Gamow who employed the WKB method to calculate the transition probability for an alpha particle 
to escape the coulomb barrier. In Fig. \ref{fig:ill} we show a schematic illustration of the process, 
showing the nuclear potential (shown as a well inside 10 fm) and various barriers on a log scale. The radius 
describes the relative distance
between the daughter nucleus (with charge $Z-2$) and the alpha-particle.
The full 
line shows the $\alpha$ particle energy without any screening and the 
dots  
indicate the path that the particle has to tunnel through. 
If we embed the radioactive 
nuclei in a metal containing a density of free electrons 
in-medium screening corrections have to be considered as well.
These in-medium effects can be described by an 
attractive screening potential (dashed line),
whose magnitude on general grounds decreases with distance 
from the nucleus with a typical screening length scale $R$.
(The amplitude of the screening potential is exaggerated for visual purposes.)
The screening potential also affects the alpha-decay energy which will be 
lowered. If we assume that the screening potential is constant
over the dimensions of the nucleus 
(this applies to the cases discussed here, as we will show 
below), then in a very good approximation the alpha-decay energy is lowered
by the value of the constant screening potential at small distances
(denoted by $\Delta U$, which is negative and is often called
the screening energy \cite{Assenbaum}).
Hence in the medium the alpha decay occurs with a shifted energy
$E_\alpha+\Delta U$
(shown by the dashed line) rather than $E_\alpha$.
(We note that this lowering of the alpha-decay energy has been neglected in
\cite{Kettner1}, and also in \cite{Liolios1} who also predicts a lowering
of in-medium nuclear decay lifetimes). 
In the medium, the alpha decay proceeds through an effective barrier
which is given by the sum of the
Coulomb barrier and the screening potential;
the effective barrier is lower than the one in the non-screened case. 
As Fig. \ref{fig:ill} shows the path through the screened barrier 
will actually be longer than through the unscreened barrier,
provided the screening potential varies over the distances of the
tunnel process. As a consequence one expects a longer alpha-decay lifetime
in the medium.
Here we have assumed that the decay energy and screening energy is simply additive. In 
general the energy shift will 
depend on the screening potential. However, for the weak screening cases 
to be studied here our assumption is justified. In fact it has been
experimentally verified 
in \cite{Kettner1} which find that nuclear resonance energies are lowered
by the screening energy if the target nuclei are embedded in a metallic medium. 
We also remark that in the extreme limit of sufficiently large screening 
(e.g. at very high densities), the alpha-decay lifetime becomes infinite as
the decaying state is shifted energetically below the decay threshold in
the medium.
Such a behavior is expected for example for pure alpha matter which will
show a phase transition to $^8$Be matter at densities in excess of a few
$10^9$ g/cm$^3$ \cite{Mueller92,Mueller94}.

\section{Basic Models and Results}
On the basis of the Debye scrrening model, it is implied in \cite{Kettner1}
that alpha-decay for $^{210}$Po and $^{226}$Ra can
be significantly reduced if the decay occurs in a metal at low temperatures.
For $^{210}$Po the experimental observation of such a reduction has already
been reported and is ascribed to Debye screening \cite{FAZ}.
We will use the two nuclei $^{210}$Po and $^{226}$Ra 
as examples to demonstrate that standard 
screening models predict very small changes of the lifetimes if the nuclei
are embedded in metals.

In the classical work on screening of nuclear reactions for 
astrophysical applications \cite{Salpeter1} the Debye-H\"uckel model was used for weak screening, with 
different corrections for electron degeneracy etc. Later works, 
including \cite{Salpeter2,Mitler1}, treated 
strong screening and also focused on different densities and temperatures. 
In the current case we are 
interested in the screening effect of electrons in metals at low temperature, 
which is a degenerate 
system. We therefore choose to work with the Thomas-Fermi model, 
which is basically the free electron gas 
model with additional assumptions on the chemical potential 
(see \cite{TFmodel} for details). But we will also estimate
the screening effects on the alpha lifetimes within the Debye model.

In both models (Thomas-Fermi and Debye) the Coulomb potential
between two charges $Z_1,Z_2$  has to be replaced 
by an effective potential of Yukawa
type $V_{\rm eff}$ if medium effects are accounted for:
\begin{equation}
V_{\rm eff}(r)= 
\frac{Z_1 Z_2 e^2}{r} \exp(-r/R)= 
\frac{Z_1 Z_2 e^2}{r} + V_{sc} (r) 
\label{scpot}
\end{equation}
where $V_{sc}$ is the in-medium screening potential. 
In the Thomas-Fermi (TF) model, the screening length scale
(screening radius) is given by
\begin{equation}
R_{TF}=\sqrt{ \frac{\epsilon_0 E_F}{3e^2\rho}}
\end{equation}
Here $\rho$ is the electron density and $E_F$ the fermi energy. 
In natural units for metals one has
\begin{equation}
R_{TF}=3.7\cdot 10^{-10} \sqrt{ \frac{E_F[eV]}{\rho[10^{22}cm^{-3}] }  }\,\, cm
\end{equation}
Using typical values
($\rho\sim 10\cdot 10^{22}$ cm$^{-3}$ and $E_F\sim 10$ eV \cite{TFmodel}), 
one finds 
$R_{TF}=3.7\cdot 10^{3}$ fm, which is 
more than two orders of magnitude larger 
than the nuclear dimensions. 
The 
screening potential should therefore not vary over the 
nuclear range, supporting our assumption of a constant 
value $\Delta U$ of the screening potential in the nuclear interior,
by which the screening energy will be shifted. 
{\footnote{Due to the singularity at $r=0$,
neither the bare Coulomb nor the effective Yukawa potential
can be the realistic solution at small nuclear radii.
To avoid problems, we use a small constant nuclear radius
as the inner turning point of the barrier.
}}

In the Debye-H\"uckel approximation \cite{Salpeter1}, 
the screening radius $R_D$ is temperature dependent. 
If one puts in typical numbers 
for metals one finds
\begin{equation}
R_D=2.18\cdot 10^{-8} \sqrt{T[K]}\,\,cm
\end{equation}
Assuming a temperature of 4 Kelvin \cite{Kettner1},  $R_D$ is 
two orders of magnitude larger than $R_{TF}$, again 
justifying our assumption of a constant screening potential inside the nucleus. However, the metal environment 
with its degeneracy requires modifications to the 
Debye-H\"uckel model (see \cite{Salpeter1,Mitler1}),
so the comparison should be regarded with some caution. 

The alpha-decay halflife is dominated by the barrier penetration probability 
$P$ which one can evaluate within
WKB theory, resulting in 
\begin{equation}
P\sim \exp(-2\,S(E))
\end{equation}
where $S(E)$ is the action integral. For a non-screened  Coulomb barrier 
one has
\begin{equation}
S(E)=\sqrt{\frac{2m}{\hbar^2}}\int_{R_{\rm nuc}}^{R_0} \sqrt{\left( 
\frac{e^2 Z_1 Z_2}{r}-E \right)     }\,dr
\label{nsc}
\end{equation}
where $R_{\rm nuc}$ is the nuclear radius 
(we will use $R_{\rm nuc}=1.22\cdot A^{1/3}$ fm, where $A$ is the nuclear mass number) 
and $R_0$ is the outer 
turning point. In the 
case of alpha 
decay one uses $Z_1=2$ and $Z_2=Z_{nuc}$, where $Z=Z_1+Z_2$
is the charge number of the decaying nucleus. 
Then the outer turning point is 
$R_0=2Z_{\rm nuc} e^2/E$. 
In the presence of the screening potential the action becomes
\begin{equation}
S_{sc}(E)=
\sqrt{\frac{2m}{\hbar^2}}\int_{R_{\rm nuc}}^{R_1} 
\sqrt{\left( V_{\rm eff}(r)-(E+\Delta U) \right)} dr =    
\sqrt{\frac{2m}{\hbar^2}}\int_{R_{\rm nuc}}^{R_1} 
\sqrt{\left( \frac{Z_1 Z_2 e^2}{r} + V_{\rm sc}(r)-(E+\Delta U) \right)     
}\,dr
\label{sc}
\end{equation}
where $R_1$ is the outer turning point for the screened barrier 
(see Fig. \ref{fig:ill}).
As discussed above the alpha-decay energy has 
to be replaced by the shifted value $E+\Delta U$. 
We fix the value of $\Delta U$ by
assuming that $V_{\rm sc}$ is constant over the nuclear dimensions,
i.e. $V_{\rm sc}(0)= \Delta U=e^2 Z_1 Z_2/R$,
were $R$ is the screening radius.
If $V_{\rm sc}(r)$ is even constant for radii $r\le R_1$, 
then the in-medium 
effects cancel and the alpha lifetime remains unchanged. Obviously modifications of the lifetime can be 
expected, if the screening potential varies over the
radius scale of the outer turning point
(e.g. at high densities were the screening length scale $R$ is small) or 
the alpha-decay energy is small. 
Estimating $R_0$ for a typical alpha decay energy of $E=5$ MeV, one finds
$R_0 \approx 51$ fm (for $Z=90$), which is much smaller than the Thomas-Fermi or
Debye screening radii estimated above. Hence changes of the alpha decay
lifetimes should be quite small. To estimate such changes we define the ratio
between the halflives in the screened $T_{\alpha,sc}$ and unscreened case
$T_\alpha$:
\begin{equation}
{\cal R} = \frac{T_{\alpha,sc}}{T_\alpha} = \frac{P}{P_{\rm sc}}
\label{ratio}
\end{equation} 
with the penetration factors for screened and unscreened case as defined above.
We have estimated this ratio for typical alpha-decay energies and various
screening radii by calculating the action integrals 
(\ref{nsc}) and (\ref{sc}) numerically. 
The results are summarized in Table \ref{ratios}.
As one can clearly see the effects are quite small for typical
screening radii in metals
(a few $10^3$ fm in the Thomas-Fermi model). 
Moreover, Thomas-Fermi and Debye screening models
both predict that the alpha-decay halflife should increase if the nucleus
is embedded in a metal. If application of the Debye model 
is allowed to describe the in-medium screening, the screening length
scale depends on temperature. But we find that even at temperatures of order
Kelvin, the changes of the halflives should be very small.
Thus our results agree with the findings of Ref. \cite{Rubinson72}
which predicts that changes of alpha lifetimes due to electronic screening
effects should be
negligibly small (see also \cite{Emery72}).

As expected on general grounds, the in-medium modifications increase the
lower the alpha decay energy (which implies a longer 
path under the barrier) and the smaller the screening radius
(allowing for changes of the screening potential over the length
of the tunnel path). The screening models predict a sizable
enhancement once the screening radius gets of the same order as the
outer turning point (of order a few times $10^2$ fm).

The alpha-decay energies of $^{210}$Po and $^{226}$Ra are 5.30 MeV and
4.78 MeV, respectively. If we assume a typical screening length scale 
for metals of $R_{TF}=3.7\cdot 10^3$ fm (Thomas-Fermi model)
or $R_D = 4 \cdot 10^5$ fm (Debye model at 4 K), 
we find that the alpha-decay lifetimes of these two nuclei should be
very slightly enlarged if embedded in a metal.
For $^{210}$Po we find $\mathcal{R}=1.009$ (Thomas-Fermi) 
and $\mathcal{R}=1.000$
(Debye), and for $^{226}$Ra
${\cal R}=1.013$ (Thomas-Fermi) and $\mathcal{R}=1.000$ (Debye). 
In turn, if experimentally a reduction of the $^{210}$Po halflife is observed
in metals \cite{FAZ}, 
then the origin of this unexpected effect should not be due
to in-medium screening as described by the Thomas-Fermi or Debye models.

In Ref. \cite{Kettner1}, the screening energy, which a $^{210}$Po
nucleus experiences in a metal, is estimated as $\Delta U=420$ keV,
based on many studies of the observed screening enhancement of the d+d 
fusion reaction cross section in metals \cite{Raiola02,Raiola04,Raiola05}.
This screening energy corresponds to a screening radius $R=562$ fm
and using this value we find that the $^{210}$Po lifetime should
be enhanced by 45\%  due to screening corrections based on the effective
potential defined in (1),

\begin{table}[htb!]
\begin{center}
\begin{tabular}{|l|c|c|c|c|c|c|}
\hline
E/R  & $1.0\cdot 10^6$ & $1.0\cdot 10^5$ & $1.0\cdot 10^4$ & $5.0\cdot 10^3$ 
& $1.0\cdot 10^3$ & $5.0\cdot 10^2$ \\
\hline
3.0 & 1.000  & 1.000  & 1.006 & 1.025 & 1.828 & 10.95 \\
\hline
4.0 & 1.000  & 1.000  & 1.003 & 1.012 & 1.342 & 3.217 \\
\hline
5.0 & 1.000  & 1.000  & 1.002 & 1.007 & 1.184 & 1.954 \\
\hline
6.0 & 1.000  & 1.000  & 1.001 & 1.004 & 1.113 & 1.530  \\
\hline 
\end{tabular}
\end{center}
\caption{Table of halflive ratios ${\cal R}$ 
as defined in equation (\ref{ratio}) for various 
choices of the unscreened alpha-decay 
energies (in MeV) and the in-medium screening radius (in fm). All 
calculations  
were done for nuclear charge $Z=90$ and mass $A=232$.}
\label{ratios}
\end{table}

\section{Conclusion}
In summary, we have studied the in-medium effects on alpha lifetimes
if the decaying nucleus is embedded in a metal. Based on standard
screening models like the Thomas-Fermi model or the Debye approach
we have given quite general arguments why the effects should be small
and have then supported these by numerical calculations. Indeed we find
negligible changes of the lifetimes for $^{210}$Po and $^{226}$Ra, 
for which recently quite sizable effects have been predicted 
and reported \cite{Kettner1,FAZ} based
on applications of a Debye model for the host metal.
Thus our calculations do not support the exciting idea that nuclear waste
can be faster disposed of if embedded in metals at low temperatures due
to significantly reduced lifetimes. We believe that the estimate
given in \cite{Kettner1}, which has born out this idea, is 
incorrect as it does not account for the in-medium modification
of the decay energy which, in first order, cancels the barrier reduction 
for the emitted alpha particle.

We like to remark that the influence of the medium 
for alpha decay is different than for
other physical processes. It is wellknown that fusion reactions
of charged particles are enhanced at low energies due to screening effects
in the plasma \cite{Salpeter1} or in the laboratory
\cite{Assenbaum,Engstler,Raiola1}. This comes about as the 
relative energy of the fusing nuclei at the outer turning point,
where the penetration process starts, is larger than
the asymptotic energy due to screening effects. Or in an alternative
picture, the fusing nuclei at a given energy have to penetrate
through a barrier which is reduced by the screening effects.
Of course, the medium also influences the alpha decay energy, i.e.
the asymptotically measured energy in the medium is not the decay energy
of the bare nuclear system, but has to be corrected for screening effects.
However, such modifications reflect the variations of the screening potential
mainly at radii larger than the outer turning point and hence do not
change the decay halflives as the effective barrier, which the alpha particle
has to penetrate, is nearly the same in the medium as in the bare
nuclei. We mention that the medium modifications for alpha decay (or the decay
by other charged particles) is similar to the resonant screening
corrections in fusion reactions where particular care has to be taken
for the corrections of the resonance energy \cite{Salpeter2,Cussons1}.

The medium corrections to alpha decay are also different to those
for beta decay or electron captures. In both cases the weak-interaction
nuclear matrix element is unchanged by the medium, however, phase space has
to be corrected as parent and daughter nuclei have charge numbers which differ
by one unit. Hence screening effects change the effective Q-value 
for beta decays or electron captures in the medium. Modifications
of beta halflives due to screening effects in metals have for example been
reported in \cite{Limata}. For electron captures
one has additionally to consider that the capture rate is proportional
to the electron probability at the nucleus which can be strongly changed
by medium effects. A prominent example is $^7$Be, which as a bare nucleus
in vacuum, has an infinite halflife, but in atoms decays by
mainly K-shell electron capture \cite{Rolfs}. In the sun, however,
the $^7$Be lifetime is noticeably increased as the electron probability
at the nucleus is smaller in the solar plasma than in an atom
\cite{Johnson92}. Experimentally medium-modifications of the
$^7$Be lifetime have been reported in \cite{Wang}.

\begin{acknowledgements}
The author would like to thank Karlheinz Langanke for useful suggestions and invaluable 
discussion. Thanks goes also to Hans Feldmeier for helpful ideas and Jim Truran for pointing out several 
nice references. The author would like to thank GSI for its hospitality during which part of this 
work was done.
\end{acknowledgements}

\end{document}